\documentstyle[12pt]{article}
\newcommand{\beq}{\begin{equation}}
\newcommand{\eeq}{\end{equation}}
\newcommand{\pr}{\partial}
\newcommand{\lam}{\lambda}
\newcommand{\sg}{\sigma}
\newcommand{\eps}{\epsilon}
\newcommand{\mn}{\mu\nu}
\newcommand{\mnl}{\mu\nu\lambda}
\newcommand{\til}{\tilde}
\newcommand{\th}{\theta}
\newcommand{\al}{\alpha}
\begin{document}
\title{Lagrangian and Hamiltonian formulations of higher order 
Chern-Simons theories}
\author{
{Sarmishtha Kumar} \thanks {e-mail: kumar@boson.bose.res.in}\\
S.N. Bose National Centre For Basic Sciences\\
Salt Lake City, Block JD, Sector III\\
Calcutta 700 098, INDIA
}
\maketitle
\bigskip
\begin{abstract}
We consider models involving the higher (third) derivative extension of the 
abelian Chern-Simons (CS) topological term in D=2+1 dimensions. The polarisation vectors in these models reveal an identical structure with the corresponding
expressions for usual models which contain, at most, quadratic structures.
We also investigate the Hamiltonian structure of these models and show how
Wigner's little group acts as gauge generator.

\end{abstract}
\newpage
{\Large\bf{Introduction}}
\\

The abelian Chern-Simons(CS) topological action represented by,\\ 
$I_{CS}=
\frac{\theta}{2}\int d^3x(\eps^{\mnl}f_{\mu}\pr_{\nu}f_{\lambda}) $ has been studied extensively in different aspects in 2+1 dimensions. Recently the higher derivative extensions of the CS action, specially the leading third derivative order
(TCS) has been considered. For abelian vector fields the action $I_{TCS}$ can be given by $I_{TCS}=\frac{1}{2m}\int d^3x \eps^{\mnl} \Box f_{\mu}\pr_{\nu}f_{\lambda}$. It remains gauge invariant, parity odd but no more topological
unlike the usual CS term, because of the metric dependence in the additional 
covariant derivative factor \cite{DJ}. Many interesting observations 
come out in coupling this TCS term to either pure Maxwell term or usual CS term or to both of these terms \cite{DJ}.

In this work we are trying to unravel the connection of such higher derivative models with some familiar models such as the Maxwell-Chern Simons, the Proca, or the Maxwell-Chern Simons-Proca models, which contain quadratic derivative terms at most in the action.

In the section I, we have calculated the polarisation vectors of the higher derivative models considered here. Then it has been shown that the form of these polarisation vectors are coming identical with that of  conventional models. To get this we have considered the Lagrangian formulation \cite{RBT}.

In section II, we adopt the Hamiltonian analysis by considering 
a particular model. In this section the problem accounted in 
quantising these higher derivative models is illustrated \cite{LSZ,GT}.

In section III, we consider the same particular model. Here 
we have discussed about the gauge transformation property of the system; 
using the Wigner's Little Group \cite{W,W1,HKS}.
This group is shown to act as a gauge generator. 

Our metric convention is $g^{\mu\nu}=(+,-,-)$; and $\eps^{012}=+1=\eps_{012}$.  
\\

{\Large\bf{Section I. Lagrangian Analysis}}
\vskip 1.0 cm

To start with, let us first consider the pure Maxwell term coupled with the 
third derivative CS term, giving the Lagrangian,
\beq
{\cal{L}}_{MTCS} = -\frac{1}{4}F_{\mu\nu}F^{\mu\nu} + \frac{1}{2\theta}
\eps_{\mu\nu\lambda} \Box f^{\mu}\pr^{\nu}f^{\lambda}
\label{a}
\eeq
where the field strength is defined as, 
$$
F_{\mu\nu} = \pr_{\mu}f_{\nu} - \pr_{\nu}f_{\mu}
\nonumber
$$
and $\theta $ has the dimensions of mass. The eq. of motion obtained from
(\ref{a}) is,
\beq
\pr_{\mu}F^{\mu\nu} + \frac{1}{\theta}\eps^{\nu\lam\sg} \Box \pr_{\lam}f_{\sg}
= 0
\label{b}
\eeq
Substituting the solutioin for the negative energy component in terms of the 
polarisation vector $\eta_\nu$,
\beq
f_{\mu}(k) = \eta_{\mu}(k)e^{ik.x}
\label{c}
\eeq
we get,
\beq
\eta^{\mu} = \frac{1}{k^2}k^{\mu}(k.\eta) - \frac{i}{\theta}\eps^{\mu\nu\lam}
k_{\nu}\eta_{\lam}
\label{d}
\eeq
Two cases are now possible; $k^2=0$ for massles modes and $k^2\neq0$ for massive modes. Consider first the massless case. We can choose the momentum $k^\mu$
propagating along the 2nd direction so that,
\beq
k^{\mu} = (1\,\, 0\,\, 1)^T ; k_{\mu} = (1\,\, 0\,\, -1)^T
\label{e}
\eeq
Replacing $k^\mu$ and $\eta^{\mu}= (\eta^{0} \eta^{1} \eta^{2})^T$ in 
(\ref{d}), we get
\beq
k.\eta = 0
\nonumber
\eeq
So that $\eta^\mu$ has the form, $\eta^{\mu}=(\eta^{2} \eta^{1}\eta^{2})^T$.
Now using the gauge invariance of the model (\ref{a}), $\eta^{\mu}$ may be further reduced to 
\beq
\eta^{\mu} = (0\,\, a\,\, 0)^T
\label{f}
\eeq
where 'a' is some arbitrary parameter. So solution in (\ref{f}) is valid, can be easily verified from Eq.( \ref{d}). 
This shows that the model in (\ref{a}) has one massless excitation \cite{ DJ}.

Now we investigate for the massive case i.e $k^2\ne0$, where we are allowed
to go to a rest frame. The momentum can be chosen as $k^{\mu }=(\Lambda,0,0)$.
Express $\eta^\mu$ in the rest frame as
$$
\eta^{\mu}(0) = (\eta^{0}(0),\eta^{1}(0),\eta^2( 0))^T
\nonumber
$$
Using these structures for $k^\mu$ and $\eta^\mu$ in Eq.(\ref{d}), we get
$$
\eta^{1}(0) = -i\frac{\Lambda}{\theta}\eta^{2}(0)
\nonumber
$$
\beq
\eta^{2}(0)=  i\frac{\Lambda}{\theta}\eta^{1}(0)
\label{g}
\eeq
Mutual consistency of the above equations yields,
\beq
\Lambda = |\theta|
\label{h}
\eeq
On the otherhand, using the gauge invariance of the model, $\eta^{0}(0)$ can 
be set equal to zero. Thus in the rest frame the required polarisation vector 
is,
$$
\eta^{0}(0) = ( 0 ,\eta^{1}(0), i\frac{|\theta|}{\theta}\eta^{1}(0) )^T
\nonumber
$$
modulo a normalisation factor. this can be fixed from the condition,
$$
\eta^{*\mu}(0)\eta_{\mu}(0) = -1
\nonumber
$$
so that $\eta^\mu$ finally takes the form,
\beq
\eta^{\mu}(0) = \frac{1}{\sqrt2}( 0\,\, 1\,\, i\frac{|\theta|}{\theta} )^T
\label{i}
\eeq
Note that $\eta^\mu$ naturally satisfies the transversality condition
$k.\eta = 0$.

We can now compare these results with those in the Maxwell case( for the massles mode) and in the Maxwell-Chern-Simons( MCS) case( for the massive mode). First recall that the 
Maxwell Lagrangian in D=2+1 dimensions,
\beq
{\cal L}_{M}=-\frac{1}{4}F^{\mn}F_{\mn}
\label{j1}
\eeq
yields a single massless mode with a polarisation vector identical to (\ref{f}).
So the massless excitation in (\ref{a}) has similar characteristics of the massless mode of the Maxwell part present in the ${\cal L}_{MTCS}$.

Similarly the MCS Lagrangian,
\beq
{\cal L}_{MCS} = -\frac{1}{4}F^{\mn}F_{\mn} - \frac{\theta}{4}\eps^{\mnl}
f_{\mu}F_{\nu\lambda}
\label{j2}
\eeq
gives a single massive mode.  The polarisation vector in the rest frame is 
given by \cite{RBT,Gir},
\beq
\eta^{\mu}(0) = \frac{1}{\sqrt2}( 0\,\,  1\,\,  i\frac{|\theta|}{\theta} )^T
\label{j3}
\eeq
which is identical  structure to (\ref{i}). Thus the massless(massive) mode
of MTCS model behaves, as far as the structure of polarisation vector is 
concerned, excatly like the corresponding modes in the Maxwell(MCS) theory.

Next we will consider the model containing both CS and TCS terms. The Lagrangian is given by,
\beq
{\cal L}_{CSTCS} = \frac{1}{4}\eps^{\mnl} \Box f_{\mu}F_{\nu\lam} +
\frac{m^2}{4}\eps^{\mnl}f_{\mu}F_{\nu\lam}
\label{k}
\eeq
where 'm' has the dimension of mass. The equation of motion following from (\ref{k})
is,
\beq
\eps^{\mnl}(\Box + m^2)F_{\nu\lam} = 0
\label{k1}
\eeq
Again substitution of Eq.(\ref{c}) in (\ref{k1}) yields,
\beq
\eps^{\mnl}(k^2 - m^2)(k_{\nu}\eta_{\lam} - k_{\lam}\eta_{\nu}) = 0
\label{k2}
\eeq
once again there can be two possibilities, massless modes for $k^2 =0$ and massive modes for $k^2\ne0$. Following the same procedure we get no physical massless excitation. So the case may be omitted. Now for the massive modes, let us take the rest frame configuration i.e $k^\mu = (\Lambda,0,0)^T$.
Now using the Eq. of motion (\ref{k2}) and the gauge invariance of the model
the polarisation vector comes out as,
\beq
\eta^{\mu} = (0\,\, \alpha\,\, \beta)^T
\label{k3}
\eeq
where $\alpha,\beta$ are some arbitrary parameters. Also the mass mode $\Lambda$
is found from the Eq. of motion as,
$$
\Lambda = |m|
\nonumber
$$
Therefore the model has a spectrum containing two modes of mass m. Such a 
spectrum has a close analogy with that of the Proca model.

The last model to be considered comprising all three terms, the Maxwell, the CS
and the TCS. The Lagrangian takes the form as,
\beq
{\cal L} = -\frac{1}{4}F^{\mn}F_{\mn} + \frac{\theta}{4}\eps^{\mnl}f_{\mu}
F_{\nu\lam} + \frac{1}{4m}\eps^{\mnl} \Box f_{\mu}F_{\nu\lam}
\label{l}
\eeq
where $\theta$ and m are the distinct mass parameters. The Eq. of motion 
following from (\ref{l}) as 
\beq
\pr_\mu F^{\mn} + \frac{\theta}{2}\eps^{\nu\alpha\lambda} F_{\alpha\lambda}
+ \frac{1}{2m}\eps^{\nu\alpha\beta} \Box  F_{\alpha\beta} = 0
\label{l1}
\eeq
As before substituting the gauge field $f_\mu$ defined in (\ref{c}) in Eq.(\ref{l1}) we get,
\beq
\eta^\nu = \frac{1}{k^2}[ k^\nu(k.\eta) + i(\theta - \frac{k^2}{m})\eps^{\nu\alpha\beta}k_{\alpha}\eta_{\beta} ]
\label{l2}
\eeq
Now following identical steps as before one finds that only the massive modes
are to be accounted. In the rest  frame configuretion $k^\mu =(\Lambda,0,0)^T$
the polarisation vector comes out as,
\beq
\eta^{\mu} = \frac{1}{\sqrt2}(0\,\,\,1\,\,\,{\mp}i\frac{|\Lambda|}{\Lambda})^T
\label{l3}
\eeq
Thus $\eta^\mu$ is not only complex but also shows a dual characterisation.
Furthermore, mutual consistency of relations akin to (\ref{g}) show that the mass of these modes are given by,
\beq
\Lambda^2 = \frac{m^2}{2}\left[\, (1+\frac{2\theta}{m}) \pm 
\sqrt{1+\frac{4\theta}{m}} \,\, \right]
\label{l4}
\eeq
Eq.(\ref{l4}) indicates four distinct possibilities for $\Lambda$. 
$$
\Lambda_1 = \frac{m}{2}\left[\,1+\sqrt{1+\frac{4\theta}{m}}\,\,\right]$$
$$\Lambda_2 = \frac{m}{2}\left[\,1-\sqrt{1+\frac{4\theta}{m}}\,\,\right]$$
$$\Lambda_3 = -\frac{m}{2}\left[\,1+\sqrt{1+\frac{4\theta}{m}}\,\,\right]$$
$$\Lambda_4 = -\frac{m}{2}\left[\,1-\sqrt{1+\frac{4\theta}{m}}\,\,\right]$$

Out of these four possibilities, depending upon the sign of $\Lambda$
two will lead to the result 
\beq
\eta^\mu = \frac{1}{\sqrt2}( 0\,\, 1\,\, +{i})^T
\label{l5}
\eeq
and the other two will give just the complex conjugate of (\ref{l5}), i.e
\beq
{\eta^\mu}^* = \frac{1}{\sqrt2}( 0\,\, 1\,\, -{i})^T
\label{l6}
\eeq

Thus only two distinct types of massive polarisations are possible.
These findings are quite close to that obtained in the Maxwell-Chern Simons-Proca model. The Lagrangian of M-CS-P can be given as,
\beq
{\cal L}_{MCSP} = -\frac{1}{4}F^{\mn}F_{\mn} + \frac{\theta}{4}\eps^{\mnl}
f_{\mu}\pr_{\nu}f_{\lam} + \frac{m^2}{2}f^{\mu}f_{\mu}
\label{m}
\eeq
where the mass modes are \cite{RK},
\beq
\Lambda_{\pm} = \sqrt{\frac{\theta^2}{4} + m^2} \pm \frac{\theta}{2}
\label{m1}
\eeq
The polarisations for these massive modes in rest frame are exactly identical
\cite{RBT} to (\ref{l5}) and (\ref{l6}).

\vskip 1.5 cm
{\Large\bf{Section II.  Hamiltonian Analysis}}
\vskip 1 cm

So far we have considered the Lagrangian formalism. Now we can discuss the 
Hamiltonian structure of such higher derivative models. Due to the presence of 
second order time derivative either the momenta or Hamiltonian will be non-trivial. So to get a better understanding let us consider only the Extended MCS model(i.e MTCS model) for simplicity.

Let us recall the Lagrangian ${\cal L}_{MTCS}$ in (\ref{a}). Now following 
Ostrogradski formalism for higher-order Lagrangian \cite{LSZ,GT}, 
 presence of the second order derivative term leads to the
following canonical momenta,

\begin{eqnarray}
p_{0} = \frac{\pr {\cal L}}{\pr {\dot{f_0}} } - \frac{d}{dt}\frac{\pr {\cal L}} {\pr {\ddot{f_0}}} = -\frac{1}{2\theta}\eps_{ij}\pr_{i}\dot{f_j}
\label{4a}
\end{eqnarray}
\begin{eqnarray}
p_{i} &=& \frac{\pr {\cal L}}{\pr \dot {f^i}} - \frac{d}{dt}\frac{\pr {\cal L}}
{\pr \ddot{f^i} }\nonumber \\
 &=& - F_{0i} - \frac{1}{2\theta}\eps_{ij}\Box f_{j} +\frac{1}{2\theta}\eps_{ij}\pr_{j}\dot{f_0} 
\label{4c}
\end{eqnarray}
\begin{eqnarray}
{\tilde p_{0}} = \frac{\pr {\cal L}}{\pr \ddot {f^0} }
= \frac{1}{2\theta}\eps_{ij}\pr_{i}f_{j}
\label{4b}
\end{eqnarray}
\begin{eqnarray}
{\tilde p_{i} } = \frac{\pr {\cal L}}{\pr\ddot{f^i}}
= \frac{1}{2\theta}\eps_{ij}F_{0j}
\label{4d}
\end{eqnarray}

where the canonically conjugate pair can be identified as [${f_{\mu},p^{\mu}}$]
and [${\dot {f_\mu},\tilde {p^\mu} }$], thus total twelve phase space variables span the space. Now we can easily identify the three primary constraints,
\beq
\Omega_{0} = p_{0} + \frac{1}{2\theta}\eps_{ij}\pr_{i}\dot{f_{j}}
\label{6a}
\eeq
\beq
\Omega_{i} = \tilde{p_i} - \frac{1}{2\theta}\eps_{ij}F_{0j}\,\,\,\, for \,\,i=1,2
\label{6b}
\eeq
\beq
\Omega_{3} = \tilde{p_0} - \frac{1}{2\theta}\eps_{ij}\pr_{i}f_{j}
\label{6c}
\eeq

The canonical Hamiltonian takes the form,,
\begin{eqnarray}
{\cal H} &=& \dot{f_\mu}p^{\mu} + \ddot{f_\mu}\tilde{p^\mu} - {\cal L} \nonumber\\ 
         &=& \dot{f_0}p_{0} + {2\theta}p_{i}\eps_{ik}\tilde{p_k} - 
            p_{k}\pr_{k}f_0 - {2\theta^2}{\tilde p_{i}}^2 -
            \tilde{p_i}{\nabla}^2f_i \nonumber \\ &+& \frac{1}{4}(F_{ij})^2 
+ \frac{1}{2\theta}\eps_{ij}{\nabla}^{2}f_{0}\pr_{i}f_{j}
\label{5}
\end{eqnarray}

Now considering the above hamiltonian, the time conservation of the primary  
 constraints lead to the secondary constraints,
\beq
\Omega_{4} = p_{0} + \pr_{i}\tilde{p_i} 
       = \Omega_{0} + \pr_{i}\Omega_{i}
\label{6d}
\eeq
\beq
\Omega_{5} = \pr_{k}p_{k} + \frac{\theta}{2}\eps_{ij}{\nabla^2}\pr_{i}f_{j}
\label{6e}
\eeq

Therefore above total set of constraints (\ref{6b})-(\ref{6e}) ( including primary as well 
as secondary) denote the independent set of constraints out of which we can    
identify $\Omega_i$'s as only second class and others are the firstclass 
constraints. With the help of these let us find how the canonical brackets 
are changing.
Due to the constraied nature of the system the Poisson brackets wil be replaced 
by Dirac brackets \cite{D}
$$
[X,Y]_{D} = \{X,Y\} - \{X,\Omega_{i}\}{C_{ij}}^{-1}\{\Omega_{j},Y\}$$
where ,$C_{ij} =[\Omega_{i},\Omega_{j}]_{PB}=\frac{1}{\theta}\eps_{ij}\delta(x,y)$. For this case $C_{ij}^{-1}=-\theta\eps_{ij}\delta(x,y)$ with $i,j=1,2$.

Now we mention only the nontrivial D'brackets,
$$
[f_{i}(x),p_{j}(y)]_{D} = -\delta_{ij}\delta(x,y)$$
$$
[\dot{f_{i}}(x),\dot{f_j}(y)]_{D} = -\theta\eps_{ij}\delta(x,y)$$
$$
[p_{0}(x),\tilde{p_i}]_{D} = -\frac{1}{4\theta}\eps_{ij}\pr_{j}\delta(x,y)$$
$$
[\til{p_i},\til{p_j}]_{D} = -\frac{1}{4\theta}\eps_{ij}\delta(x,y)$$
\beq
[\dot{f_i},\til{p_j}] = -\frac{1}{2}\delta_{ij}\delta(x,y)
\label{7b}
\eeq

Using these brackets one can easily show that it correctly reproduce the Euler-Lagrange Eq. of motion and also the relevant constraints from the Hamilton's Eq.
of motion. So the equivalence between the Lagrangian and Hamiltonian formalism
is satisfied. It might be observed that the algebra of $\dot {f_i}$ is 
identical to the algebra of the basic field in the usual self dual model,
$$
{\cal L}_{SD} = \frac{1}{2}f^{\mu}f_{\mu} - \frac{1}{2\theta}\eps^{\mnl}
f_{\mu}\pr_{\nu}f_{\lambda}
\nonumber
$$
\\

{\Large\bf{Section III. Wigner's Little Group and Gauge Transformation}}
\\

Next we will discuss briefly the gauge symmetries of the model concerned. It 
is clear from the model (\ref{a}), that the Extended CS term involves metric
dependence, so that it becomes non-topological. Again it has been  
already shown that there 
exist number of first class constraints which implies this is a gauge model. 
So the Extended MCS model stands for a higher derivative massive gauge theory.
Now  let us try to find out the exact gauge variation in the model using the 
concept of Wigner's little group. 
Wigner's little group E(2), which is a subgroup of the Lorentz group $SO(1,3)$
preserves the four momentum invariant, but the polarisation vector $\eta^\mu$
undergoes the gauge transformation,
$$
\eta^\mu(k) \rightarrow \eta^{'\mu}(k)= \eta^{\mu}(k)+f(k)k^\mu
$$
where $f(k)$ can be identified as the gauge parameter. It has been shown earlier \cite{W1,HKS} that translation like 
generators of Wigner's little group E(2) can act as generators of 
gauge transformation in the pure Maxwell theory in (3+1) dimensions 
where only massless quanta is admitted. Recently it is shown \cite{RBT} 
that such little group generator 
can generate gauge transformations in topologically massive gauge theories
like $B{\wedge}F$ theory in (3+1)dimensions as well as in  
Maxwell Chern-simons theory which is topological and permits 
topologically  massive quanta in (2+1) dimensions. So our natural 
question arises whether this same little 
group can provide the gauge transformations regarding the non-topological 
extended MCS model where both massless and massive modes are present.

Let us first recall   the polarisation vector for massive excitations. So the 
rest frame configuration is available, i.e $k^\mu = (|\theta|\,\,0\,\,0)^T$
In particular, for simplicity let us take $\theta>0$.
Following little group representations as in \cite{RBT} it is now 
straight forword to show that, $\eta^\mu$ undergoes the tranformation,
\begin{eqnarray}
\eta^{\mu^\prime} &=& W^{\mu}_{\nu}\eta^\nu  
= \frac{1}{\sqrt2}\left(\begin{array}{ccc} {1}&{\alpha}&{-i\alpha}\\
{0}&{1}&{0}\\ {0}&{0}&{1}  \end{array}\right) \left(\begin{array}{c}
{0}\\{1}\\{i\frac{|\theta|}{\theta} } \end{array} \right) \nonumber \\
&=& \frac{1}{\sqrt2} \left( \begin{array}{c} {\al + \al\frac{|\th|}{\th} }\\
{1}\\ {i\frac{|\th|}{\th} } \end{array} \right) \nonumber \\
&=& \frac{1}{\sqrt2}\left( \begin{array}{c} {0}\\{1}\\{i\frac{|\th}{\th}}
\end{array}\right) + (\frac{\al}{|\th|} +\frac{\al}{\th}) \left(\begin{array}{c}
{|\th|}\\  {0}\\ {0} \end{array} \right)
\nonumber\\
&=& \eta^{\mu} + \frac{2\al}{\theta}k^{\mu}
\label{w1}
\end{eqnarray}

Therefore we get  the gauge transformation on the polarisation 
vector of the massive extended MCS quanta in its rest frame \cite{W1}
where $\al$ is the gauge parameter.
It is interesting to note that the same little group representation given by
$W^{\mu}_{\nu}$ can generate gauge transformations in both MCS and Extended 
MCS cases(provided $\th$ is positive). This is due to the fact that the expression of the polarisation vector for the two cases are similar. 

On the other hand if we consider the polarisation vector $\eta^{\mu}(k)$ (since
rest frame is not available) of the massless mode of the MTCS model and succsessively operate $\eta^{\mu}(k)$ by the relevant little group $W^{\mu}_{\nu}$ which plays the same role of a gauge generator of gauge transformation in Pure Maxwell theory it can be easily seen that gauge transformation obtained  eaxctly 
matches with that of Maxwell case. Reason is same as for the massive case mentioned earlier.
\\

{\Large\bf{Section IV. Conclusion}}
\\

We have so far considered the third derivative extension of the 
topological Chern-Simons term either coupled with the Maxwell term(MTCS) 
or the Chern-Simons term (CS-TCS) itself or with both of these 
terms(MTCS-CS). It is shown however, that the polarisation 
vectors in the above three models coming out from the Lagrangian 
formulations, reveal some distinct similarity with those familiar models 
e.g the Maxwell-ChernSimons, the Proca, or the Maxwell-ChernSimons-Proca 
theories with respect to the various modes of propagation.
Now the structure of the polarisation vectors is known to yield the 
spin of the various modes in the usual models \cite{RBT,Gir}. This 
information, coupled with the mapping found here, enables one to 
predict the spin of the modes in these higher derivative models.
\\

Next we discussed the Hamiltonian formulation. We have 
considered only the MTCS model for conveniance. Due to the presence of third
order time derivative it becomes very non-trivial to get the canonical pairs
specially the momenta. So we adopted the Ostrogradski formalism for 
higher order lagrangians and successively constructed the momentum 
as well as the Hamiltonian. Here we elaborately discussed the constrained
feature of the model and computed the Dirac brackets as well.  
\\

At the end we have invesigated the gauge transformation property of that
previously mentioned model(i.e MTCS) using the Wigner's Little Group. 
It is observed that the identical representation of the Wigner's Little 
Group which causes  gauge transformation for the MCS case(which allows only the topologally massive quanta) is also able to produce gauge transformation
for the non-topologiacl massive quanta of the MTCS model.
\\

As a future prospect it might be usuful to consider a doublet of these 
higher derivative models. For second derivative (usual) models,
such doublets yield interesting consequences \cite{RK}.
\\

I am grateful to Prof. Rabin Banerjee for  suggesting this problem and also for help with valuable discussions throughout.


\newpage  
\begin{thebibliography}{99}
\bibitem{DJ} S.Deser,R.Jackiw Phy.Lett.B 451 (1999) 73
\bibitem{RBT} R.Banerjee,B.Chkraborty and Tomy Scaria, Modern.Phys.LettA 16
(2001) 853;IJMPA 16(2001)3967;R.Banerjee and B.Chkraborty Phy.Lett.B
502(2001)291.
\bibitem{LSZ}J.Lukierski, P.Stichel,W.Zakrzewski, Ann.Phys 260 (1997) 224
\bibitem{GT} D.M Gitman and I.V Tyutin,{\it{ Quantization of Fields with
Constraints}}, Springer Verlag 1990
\bibitem{W}S.Weinberg, {\it{Quantum Theory of Fields}},
 Vol 1(Cambridge University Press, 1996)
\bibitem{W1}S.Weinberg,Phys.Rev.B134(1964)882
\bibitem{HKS}D.Han,Y.S.Kim and D.Son Phys.Rev.D26(1982)3717;Phys.Rev.D
31(1985)328.
\bibitem{Gir}F.PDevecchi,M.Fleck,H.O.Girotti,M.Gomes and A.J.da Silva
Ann.Phys.(N.Y.)242(1995)275.
\bibitem{D}P.A.M.Dirac,{\it{Lectures on Quantum Mechanics}},Yeshiva
University NY, 1964.
\bibitem{RK} R.Banerjee and S.Kumar,Phys.Rev.D60(1999)085005;Phys.Rev.D 
63(2000)125008

\end{thebibliography}
\end{document}